\documentstyle[twocolumn,aps,prl,times,epsfig]{revtex}
\begin{document}
\draft
\title{Forecasting confined spatiotemporal chaos with
genetic algorithms}
\author{ Crist\'obal L\'opez$^{1}$, Alberto \'Alvarez$^{2}$ and
Emilio Hern\'andez-Garc\'\i a $^{1}$
}
\address{$^{1}$Instituto Mediterr\'aneo de Estudios Avanzados, IMEDEA
(CSIC-Universitat de les Illes Balears), 07071 Palma de Mallorca,
Spain
\\$^{2}$
SACLANT Undersea Research Centre, 19138 San Bartolomeo, La Spezia,
Italy
}
\date{March 22, 2000}
\maketitle
\begin{abstract}
A technique to forecast spatiotemporal time series is presented.
it uses a Proper Ortogonal or Karhunen-Lo\`{e}ve Decomposition to
encode large spatiotemporal data sets in a few time-series, and
Genetic Algorithms to efficiently extract dynamical rules from the
data. The method works very well for confined systems displaying
spatiotemporal chaos, as exemplified here by forecasting the
evolution of the onedimensional complex Ginzburg-Landau equation
in a finite domain.
\end{abstract}

%\begin{twocolumns}
Nonlinear time-series analysis provides tools to identify
dynamical systems from measured data \cite{books}. The approach
has been greatly developed in the last years as a powerful
alternative to linear stochastic methods in the modeling of
irregular time-series and provides, under the assumption of
deterministic behavior, useful recipes for system control, noise
reduction, and forecasting. Applications of these techniques to
situations of spatiotemporal chaos, however, is still in its
beginnings \cite{Parlitz,Orstavik}. There are two main reasons for
this: a) the large attractor dimensions of spatiotemporally
chaotic systems, increasing with system size, poses serious
difficulties to the standard methods of delay embedding and
attractor reconstruction; b) the right choice of variables is far
from obvious: whereas the time evolution of an observable at a
particular space point could be enough in some particular
situations, decaying space correlations, and propagation phenomena
turn this to be a poorly performing choice in most cases.

A very efficient method for time-series prediction using Genetic
Algorithms (GA) has been recently proposed in \cite{Szipiro} for
nonextended systems.
Comparatively small data sets are enough to use this technique,
which makes it competitive in facing difficulty a) i.e.,
prediction in the presence of attractors of large dimension. In
some cases, even non-trivial functional forms of dynamical systems
generating the data can be unveiled \cite{Yadavalli}. In this
Letter we extend the GA approach to the forecasting of {\sl
confined spatiotemporal chaos}. By this we mean the situation in
which chaotic dynamics in an extended system is strongly affected
by the presence of boundaries. Our interest in this situation,
somehow intermediate between low-dimensional chaos and homogeneous
extensive chaos, arises from its relevance to real experimental
situations\cite{exp1,exp2}, and from recent work \cite{averages}
leading to theoretical understanding: the boundaries break
translational symmetry and the resulting phase rigidity restricts
the shape of the chaotic fluctuations allowed. This manifests for
example in the appearance of nontrivial average patterns
\cite{exp1,averages} and in inhomogeneities in other statistical
characteristics \cite{exp2,Meixner}. Under these circumstances the
Empirical Orthogonal Functions (EOFs) \cite{Lumley,Sirovich}
obtained from a Proper Ortogonal Decomposition (POD, also known as
Karhunen-Lo\`{e}ve decomposition) provide an excellent basis for
describing the system dynamics. They are different from
simple Fourier modes and contain information (optimal in a precise
sense) on the broken traslational symmetry. The amplitudes of the
most important EOFs will be the variables chosen in response to
difficulty b).

We now describe more in detail our method for spatiotemporal
forecasting, in which the POD is used to encode the large
spatiotemporal data set in a few time-series, and the GA approach
is used to obtain the corresponding forecasts. Given a time series
of spatial patterns $U({\bf x},n)$, where $n=1,...,N$ labels the
temporal sequence and ${\bf x}$ the $M$ spatial points in a
$d$-dimensional mesh, the POD decomposes the fluctuations around
the temporal mean $u({\bf x},n) \equiv U({\bf x},n)-\left< U({\bf
x},n) \right>_n$ into modes ranked by their temporal variance. As
a result, a set of spatial EOFs and associated temporal amplitude
functions are obtained. The EOFs $\phi_i({\bf x})$ ($i=1,...,M$)
are the (orthogonal) eigenfunctions of the covariance matrix of
the data $C({\bf x},{\bf x'})=\left< u({\bf x},n)u({\bf x'},n)
\right>_n$ and are the spatial structures statistically more
representative of the fluctuations in the data set. Temporal
amplitude functions $a_i(n)$, describing the dynamics of the
system, are obtained from the modal decomposition $u({\bf
x},n)=\sum_{i=1}^M a_i(n) \phi_i({\bf x})$. If only $K<M$ of the
EOFs (the ones containing the highest temporal variance as
measured by the corresponding eigenvalues) are used in the
reconstruction process, the set of reconstructed patterns
\begin{equation}
u^K ({\bf x},n)=\sum_{i=1}^K a_i(n) \phi_i({\bf x})
\label{reconstruction}
\end{equation}
is still the best approximation one can obtain by linearly
combining $K$ arbitrary spatial patterns multiplied by $K$
arbitrary amplitude functions\cite{Lumley}. Even more, it has been
shown for several chaotic and even turbulent confined systems
\cite{Lumley,Sirovich} that taking a few dominating modes $K<<M$
provides a good approximation to the complete data set.

Forecasting of the amplitude functions is performed with a Genetic
Algorithm. In general, GA's are computational methods to solve
optimization problems in which the optimal solution is searched
iteratively with steps inspired in the Darwinian processes of
natural selection and survival of the fittest \cite{Holland}. Here
the optimization problem to be solved is finding the empirical
model best describing the data, that is, finding the optimum
function $F_i$ that minimizes the difference $E_i^2
\equiv \sum_{n=1}^N \left( a_i(n)-\widetilde{a_i}(n) \right)^2$ between the
values $a_i(n)$ of each time series and the corresponding
estimator given by
\begin{equation}
\widetilde{a_i}(n)=F_i\left[a_i(n-1),a_i(n-2),...,a_i(n-D)\right]\ \ ,
\label{principal}
\end{equation}
with $D+1\leq n \leq N$. Finding ${F_i, i=1,...,M}$ amounts to
identify the dynamical system behind the data set. Once found,
Eq.~(\ref{principal}) can be used to predict the future evolution
of the system. If $D$ is large enough, the existence of the exact
$F_i$'s is guaranteed by Takens theorem and its extensions
\cite{books}, but a smaller $D$ can give approximate dynamics
$F_i$ with already a reasonably low error $E_i$. In addition, we
are not looking for all the $M$ estimators but only for the $K$
associated to the dominant EOFs. In our approach, the time-series
associated to each EOF are modeled independently. More general
multivariate estimators, with each $\widetilde{a_i}$ possibly
dependent on different $a_j$'s, may in principle be used, but we
restrict to the choice (\ref{principal}) for algorithmic
simplicity.

The power of the GA resides in that a huge functional space is
explored in order to find an optimal $F_i$. Each possible $F_i$ is
a formula consisting in a combination of numerical constants,
variables, and arithmetic operators. This combination is stored in
the computer as a symbolic string. The only limitation to the
allowed functional forms (besides the limitation to arithmetic
operations) is the maximum allowed length of the symbolic string.
The search procedure begins by randomly generating an initial
population of potential estimators $F_i$ that will be subjected to
the evolutionary process. The evolution is carried out by
selecting from the initial population the strongest individuals,
i.e. the functions that best fit the data, giving a smaller $E_i$.
In practice, only a temporal part of the data set is used in this
step (the {\sl training set}), whereas the rest of the data are
used later for validating the efficiency of the prediction method
({\sl validating set}). The strongest strings choose a mate for
reproduction while the weaker strings disappear. `Reproduction'
consists in interchanging parts of the symbolic strings (the
`genetic material') between the two mating individuals. As a
result, a new generation of individuals (which includes the
original `parent' string) is generated. The new population is then
subjected to mutation processes that change, with low probability,
small parts of the symbolic strings. The evolutionary steps are
repeated with the new generation, and the process is iterated
until an optimum individual is finally found or after a fixed
number of generations. Further details about the implementation of
the algorithm can be consulted in \cite{AlJCP}.

The formulae $F_i$ are only optimized for predicting the value of
$a_i(n)$ in terms of the $D$ amplitudes immediately before in
time. We call this `one-step-ahead forecast'. One can in principle
iterate the formulae to obtain successively predictions for
$a_i(n+1)$, $a_i(n+2)$, etc. But this will normally lead to
results rapidly diverging with respect to the correct values
because of error accumulation and amplification \cite{Szipiro}.

However, GA's can
be designed specifically to forecast values of the time series not
necessarily in the immediate future. For example, finding the
function $F_i^T$ minimizing the error between the actual series
and the estimator
\begin{equation}
\widetilde{a_i}^T(n)=F_i^T\left[ a_i(n-T),a_i(n-T-1),...,a_i(n-D)
\right],
\label{principal1}
\end{equation}
with $D+1\leq n \leq N$, allows direct prediction of $a_i(N+T)$,
that is prediction $T$-steps ahead, without iteration.

\begin{figure}
\epsfig{file=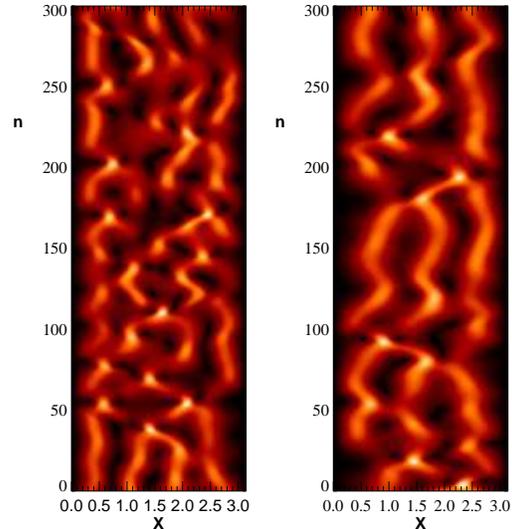,width=.8\linewidth} \caption{Spatiotemporal
evolutions of $U(x,n)$, as given by the CGLE for $q=0.12$ (left)
and $q=0.16$ (right). Black corresponds to $U=0$, and lighter gray
to high values of $U$.}
\label{fig:patterns}
\end{figure}

{\it Numerical results}. To illustrate the forecasting method we
generate a data set from numerical simulation of a well-studied
model equation displaying spatiotemporal chaos, the
one-dimensional Complex Ginzburg-Landau equation (CGLE) ,
supplemented with Dirichlet boundary conditions at the ends of a
finite interval\cite{Sirovich}. It is convenient for our purposes
to write it as
\begin{equation}
\partial_t A(x,t) = q^{2}(1+\alpha)\partial_x^{2} A
+ A -(1+i\beta)A|A|^{2},
\label{CGL}
\end{equation}
%\begin{equation}
%\frac{\partial A(x,t)}{\partial t}= q^{2}(i+c_0)
%\frac{\partial^{2} A(x,t)}{\partial x^{2}}
%+\rho A +(i-\rho)A|A|^{2},
%\end{equation}
where $q$, $\alpha$, and $\beta$ are real and positive and
$A(x,t)$ is a complex-valued field. We solve it in the interval
$[0,\pi]$ so that the boundary conditions read $A(0)=A(\pi)=0$. By
simple scaling of the spatial coordinate one sees that this is
equivalent to rewriting the equation with $q=1$, but solving it in
a domain of size $L=\pi/q$. Thus the parameter $q$ is equivalent
to an inverse system size, and decreasing it is equivalent to
increasing system size. Following \cite{Sirovich} we fix
$\alpha=4$ and $\beta=-4$ \cite{parameters}. For $q < 0.2 $ the
system displays spatiotemporal chaos for most of the initial
conditions. Decreasing $q$ one encounters the regime of confined
spatiotemporal chaos we are interested in before approaching
homogeneous extensive chaos at large system sizes ($q \rightarrow
0$) \cite{largesystem}. According to \cite{Sirovich}, the
correlation dimension of the dynamical attractor for $q=0.14$ is
$9.08$. We sample our simulation every $\tau=0.1$ time units and
at spatial locations separated $\Delta=\pi/100$ space units, and
follow it for $80$ time units ($800$ samples) after discarding the
initial transient starting from random initial conditions (this
sampling leads to $N=800$ and $M=100$). This will be our `training
set' to be feed into the GA. The simulation is then continued for
a few more time units, to provide the `validation set' which is
hidden to the GA. It is used later to check the accuracy of the
predictions.

\begin{figure}
\epsfig{file=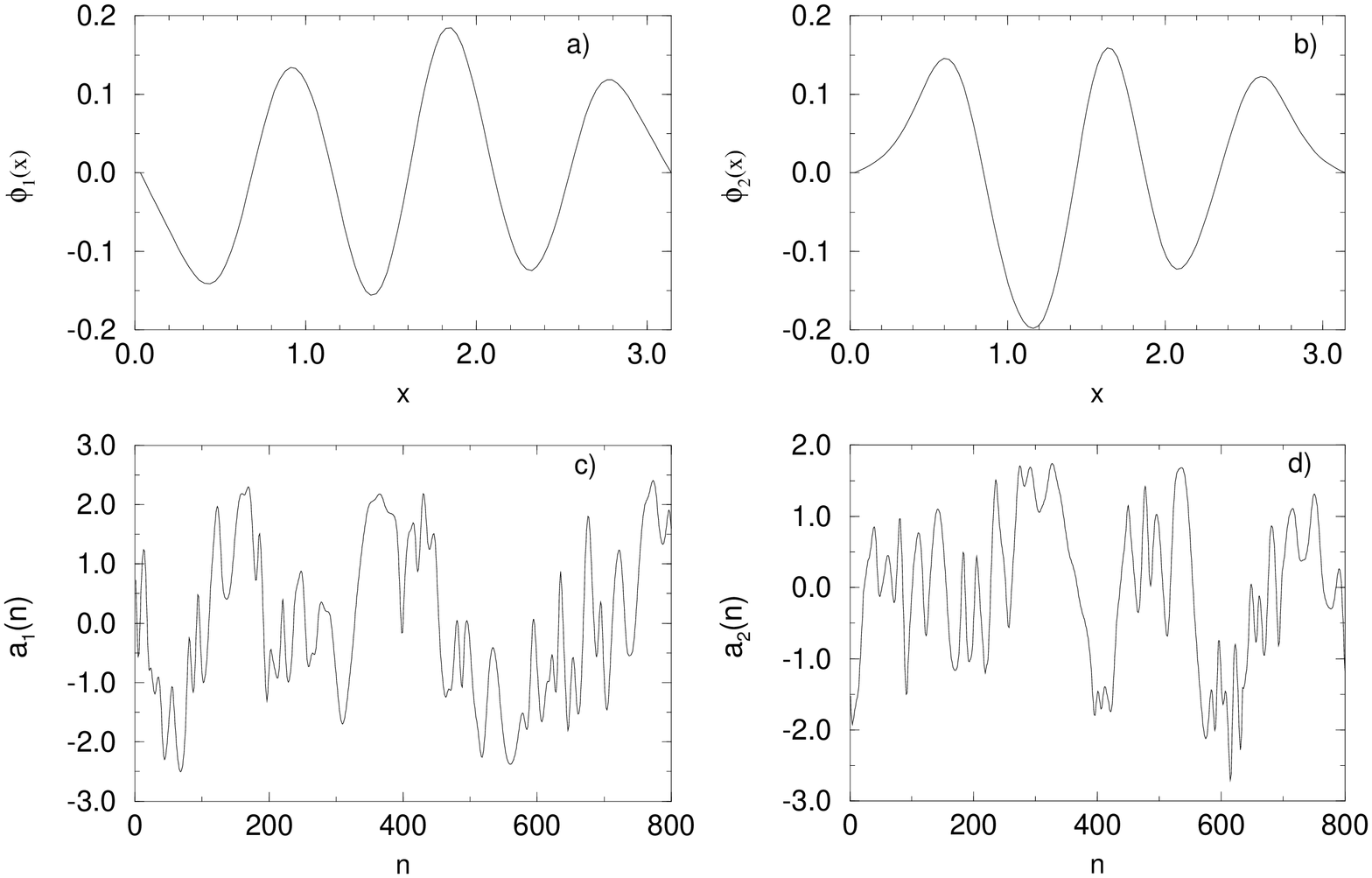,width=.6\linewidth,angle=270} \caption{The first
two EOFs (a and b), and the corresponding time amplitude functions
(c and d) from the training set at $q=0.16$.}
\label{fig:decomposition}
\end{figure}

We choose as the basic field to be forecasted the {\sl modulus}
$U(x,n)=\left|A\left(x,t=n\tau\right)\right|$ of the complex
field. The algorithm seems to perform slightly better in
forecasting the real or the imaginary parts of $A$, but we use $U$
to show that the algorithm works well with nonlinear combinations
of the basic dynamical quantities. In Fig.~(\ref{fig:patterns}) we
show parts of typical spatiotemporal evolutions for $q=0.12$ and
$q=0.16$. Clearly, reducing $q$ decreases the spatial scales, as
corresponding to an effectively larger system size, but also the
complexity of the evolution is increased. In both cases it is
clear that the motion of the dynamical structures is constrained
by the presence of the walls, as corresponding to {\sl confined}
spatiotemporal chaos.

We solve Eq.~(\ref{CGL}) for $q=0.18,0.16,0.14,0.12$ and perform
the POD on the fluctuations $u(x,n)$ of the modulus around its
temporal mean value in the resulting data sets. The number of
relevant EOFs (which we define to be those accounting for at least
$99\%$ of the data variance\cite{Sirovich}) are respectively $9$,
$11$, $13$, and $15$. We note that this confirms the expected
approximate linear scaling of the number of EOFs with increasing
system size $L$ ($\propto q^{-1}$)\cite{Zoldi}. It is somehow
surprising that this extensive scaling appears even when chaos is
not homogeneous, but still influenced by the boundaries. This fact
has been observed in other systems before \cite{exp2,Meixner}. For
illustrative purposes, we show in Fig.~(\ref{fig:decomposition})
the two most relevant EOFs from our training set at $q=0.16$, and
the corresponding temporal amplitude functions. The chaotic
character of these series is evident. 

\begin{figure}
\epsfig{file=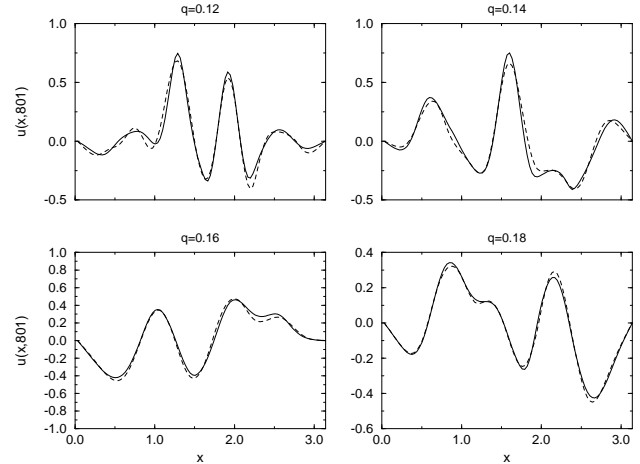,width=.75\linewidth,angle=270}
\caption{The forecasted moduli fields (dashed line)
as compared to the real ones (solid) for one-step-ahead prediction
for several values of $q$.}
\label{fig:forecasting}
\end{figure}

We next apply the GA to each of the amplitude functions of the
relevant EOFs. We use the following parameters for all the values
of $q$: number of generations in the evolutionary process $2000$,
number of individuals in each generation $120$, maximum number of
symbols allowed for each symbolic string $20$, number of delays in
(\ref{principal}) or (\ref{principal1}) $D=18$. Tuning of these
parameters for each particular value of $q$ would improve
forecasting, but would make comparisons more difficult.
Predictions for the field $u(x,t)$ are then build up by
reconstruction according to (\ref{reconstruction}) with $K$ the
number of relevant EOFs defined above. In
Fig.~(\ref{fig:forecasting}) we show the one-step-ahead forecasted
fields, more concretely the prediction for the first step beyond
the training set, $n=801$. It is compared with the actual
numerical pattern in the validation set, for $q=0.12, 0.14 $ and
$q=0.16$, displaying an excellent performance.

\begin{figure}
\epsfig{file=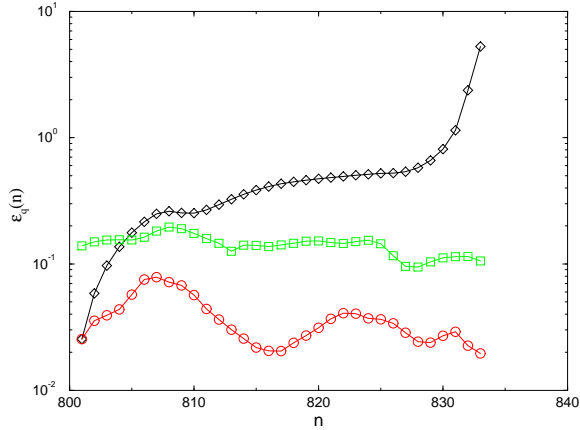,width=.7\linewidth,angle=270} \caption{Errors
as a function of $n$ in the validation set, for $q=0.16$. Circles:
one-step-ahead prediction. Diamonds: iteration of the
one-step-ahead formulae starting from the training set ($n \leq
800$). Squares: five-steps-ahead prediction. }
\label{fig:errn}
\end{figure}

We quantify the quality of the prediction in terms of the mean
square error $\epsilon_q(n)$:
\begin{equation}
\epsilon_q^{2}(n) \equiv \frac{1}{M} \sum_{j=1}^{M}
\left( \widetilde{u}^K(x=j\Delta,n) - u(x=j\Delta,n) \right)^{2},
\end{equation}
where $\widetilde{u}^K(x,n)$ is the predicted pattern
reconstructed from Eq.~(\ref{reconstruction}) and $u(x,n)$ is the
actual pattern from the validation set. As stated before, GA's can
be used to predict future values some time steps ahead, without
the need of iterating the one-step-ahead predictor (which early
becomes useless because of the expected exponential growth of
errors). Figure (\ref{fig:errn}) shows $\epsilon_{q}(n)$ as a
function of $n$ for $q=0.16$ calculated from: a) the
one-step-ahead prediction formulae obtained from the training set,
but applied to obtain the pattern at step $n$ from the previous
$D$ values in the validation set; b) iteration of the
one-step-ahead formulae starting from the last $D$ data in the
training set; c) five-steps-ahead prediction from a formula of the
type (\ref{principal1}) with $T=5$, obtained by the GA in the
training set, and used into the validation set. We see that the
improvement in accuracy is notorious when iteration is avoided. We
note that the errors in methods a) and c) remain bounded even when
$n$ is far from the values from which the prediction formulas were
estimated (i.e. the training set $n<800$). This confirms that the
method is not simply fitting data, but rather it has really found
approximate dynamical rules within the deterministic
spatiotemporal series.

\begin{figure}
\epsfig{file=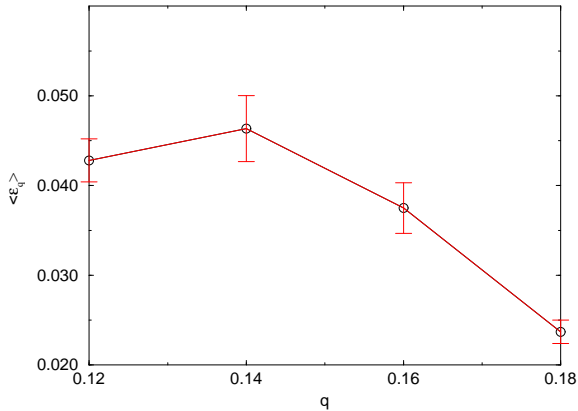,width=.7\linewidth,angle=270}
\caption{Mean
error for one-step-ahead prediction in the validation set as a
function of $q$.}
\label{fig:errq}
\end{figure}

Figure (\ref{fig:errq}) displays the average error
$<\epsilon_q>$, which is the temporal average of
$\epsilon_q(n)$ with $n$ in the validation range displayed in
Fig.~(\ref{fig:errn}), as a function of $q$ (for one-step-ahead
prediction). Despite we are including more EOFs in the
reconstruction for decreasing $q$, the prediction error shows
a tendency to increase.
This is a consequence of the increase in complexity (and in
attractor dimension) of the dynamics by the effective increase in
system size ($\approx q^{-1}$). Since we keep the number of delays
$D$ fixed, the embedding of the data set becomes more incomplete
at smaller $q$ and the prediction deteriorates. In addition, for
smaller $q$ the {\sl confined} or {\sl boundary influenced}
character of the spatiotemporal chaos in the system is lost and a
description in terms of local structures will be certainly more
efficient \cite{Parlitz}.

In summary, we have presented a method to forecast the evolution
of spatially extended systems based in the combination of POD and
GA's. The method performs very well in situations of confined
spatiotemporal chaos as exemplified by the CGLE in a finite
interval. We mention here that we are exploring the possibilities
of the method for prediction from noisy natural data sets. Results
obtained in forecasting Sea Surface Temperature patterns in an
area of the Mediterranean Sea \cite{GRL} are encouraging.

We acknowledge financial support from CICYT (MAR98-0840) and DGICYT (PB94-1167).

%\end{twocolumns}

\end{document}